\documentclass[letter]{aa}  

\usepackage{graphicx}
\usepackage{txfonts}
\usepackage{hyperref}
\hypersetup{
    colorlinks=true,
    linkcolor=cyan,
    citecolor=cyan,
    urlcolor=cyan,
    }

\newcommand{\orcid}[1]{\protect\href{https://orcid.org/#1}{\protect\includegraphics[width=8pt]{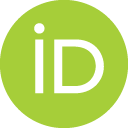}}}

\begin{document}

   \title{V4141 Sgr: Outflows and repeated outbursts}

   \author{Jaroslav Merc\orcid{0000-0001-6355-2468}
          \inst{1,2}\thanks{\email{jaroslav.merc@mff.cuni.cz}}
          \and
          Joanna Miko\l{}ajewska\orcid{0000-0003-3457-0020}
          \inst{3}
          \and
          Thomas Petit\inst{4}
          \and
          Berto Monard \inst{5}
          \and
          Stéphane Charbonnel\orcid{0009-0006-0817-4699}\inst{4}
          \and
          Olivier~Garde\orcid{0000-0002-7850-8360}\inst{4}
          \and
          Pascal~Le~Dû\orcid{0000-0003-2385-0967}\inst{4}
          \and
          Lionel Mulato\orcid{0000-0002-6822-9368}\inst{4}
          \and
        Tadashi Kojima\inst{6}
          }

    \authorrunning{Jaroslav Merc et al.}

   \institute{
             Astronomical Institute of Charles University, V Hole\v{s}ovi\v{c}k{\'a}ch 2, Prague, 18000, Czech Republic
             \and
             Instituto de Astrof\'isica de Canarias, Calle Vía Láctea, s/n, E-38205 La Laguna, Tenerife, Spain
             \and
             Nicolaus Copernicus Astronomical Center, Polish Academy of Sciences, Bartycka 18, 00–716 Warsaw, Poland
             \and
             Southern Spectroscopic Project Observatory Team (2SPOT), 45, Chemin du Lac 38690 Châbons, France
             \and
             Kleinkaroo Observatory, Sint Helena 1B, PO Box 281, Calitzdorp 6660, South Africa
             \and
             Tsumagoi, Gunma, Japan
             }

   \date{Received March 24, 2025; accepted April 23, 2025}

 
  \abstract
   {In this work, we analyze the ongoing brightening of the poorly studied symbiotic star V4141~Sgr and examine its long-term variability. We present new low-resolution spectroscopic observations of the system in its bright state and combine them with multi-color photometric data from our observations, ASAS-SN, ATLAS, and \textit{Gaia} DR3. To investigate its long-term evolution, we also incorporate historical data, including photographic plates, constructing a light curve spanning more than a century. Our analysis reveals that V4141~Sgr has undergone multiple outbursts, with at least one exhibiting characteristics typical of "slow" symbiotic novae. The current outburst is characterized by the ejection of optically thick material and possibly bipolar jets, a phenomenon observed in only a small fraction of symbiotic stars. These findings establish V4141~Sgr as an intriguing target for continued monitoring.
}

   \keywords{binaries: symbiotic --
                Stars: winds, outflows --
                Stars: individual: V4141~Sgr
               }

   \maketitle
%

\section{Introduction}
V4141~Sgr (=Th 4-4) is a peculiar symbiotic star that has been studied far less than it deserves. Initially classified as a planetary nebula \citep[PN;][]{1964CoBos..28....1T,1967cgpn.book.....P}, it was later also described as a peculiar Be star \citep{1975TrAlm..25...23K} or a low-excitation, high-density PN \citep[e.g.,][]{1989SvAL...15...13K}. Its possible symbiotic nature was first suggested by \citet{1984PASA....5..369A}, who reported the detection of a late M-type continuum, strong H$\alpha$, \ion{He}{i}, and possibly \ion{He}{ii} emission. A mid-M spectral type was also proposed by \citet{1983RMxAA...8...39M}. Further studies identified additional strong emission lines, such as [\ion{O}{iii}] and possibly faint Raman-scattered \ion{O}{vi} \citep{1987A&AS...68...51S}\footnote{The same spectrum was also analyzed by \citet{1997A&A...327..191M} and is discussed in the Appendix \ref{app:1984}. We cannot confirm the detection of \ion{O}{vi} lines.}, although other authors did not confirm the detection of the latter since then. \citet{1992PASP..104.1187G} did not observe well-defined absorption bands and noted that the spectrum resembled a low-excitation, high-density PN, but their diagnostic based on [\ion{O}{iii}] line ratios suggested an infrared S-type ("stellar") symbiotic system classification rather than a PN. Interestingly, \citet{1997A&A...327..191M} found that the \ion{He}{i} line ratios and IRAS colors were more consistent with a D-type ("dusty") symbiotic classification. Building on the aforementioned and an unpublished K-band spectrum showing strong CO 2.3 $\mu$m absorption bands, indicative of an M6 giant companion (see the Appendix \ref{app:ir}), \citet{2000A&AS..146..407B} included V4141~Sgr among the confirmed symbiotic binaries. The object has since been included as such also in later symbiotic catalogs \citep{2019ApJS..240...21A,2019RNAAS...3...28M,2019AN....340..598M}.

Until very recently, the only dedicated study of V4141~Sgr in the past 25 years was that of \citet{2001A&A...376..978K}, who analyzed its photometric and spectroscopic evolution from 1970 to 2001, building on their earlier work \citep{1989SvAL...15...13K}. Their study described an apparent transformation from a Be star into a PN within three decades. They observed dramatic changes in the continuum shape, with a strong blue continuum present in the early 1970s that later disappeared, significant variations in the intensity and presence of emission lines (interpreted as an increase in the effective temperature of the central star), and long-term photometric variability. The light curve showed a decline in brightness by at least 2.5 magnitudes between 1970 and 1990. Additionally, the authors claimed that the archival data from the Palomar Survey suggested that the object had already been bright for at least a decade before their systematic monitoring began. 

Despite briefly mentioning literature reports of an M-type continuum and a possible connection to 'slow' symbiotic novae, \citet{2001A&A...376..978K} ultimately concluded that the star could be classified “without any doubt” as a PN. More recently, \citet{2025Galax..13....5K} analyzed additional observations, extending previous studies with data up to 2024 showing no drastic changes since 1990. Their findings acknowledge the possibility that V4141~Sgr is a symbiotic star, though it appears as a low-excitation PN.

Apart from these works, V4141~Sgr received little attention until we detected its brightening in early February 2025. In this paper, we present spectroscopic follow-up observations of this recent outburst, revealing optically thick outflows from the system. We also analyze its recent and long-term variability, demonstrating that, in addition to the outburst observed from 1970 (or 1960) to 1990, the system underwent at least one additional brightening in the 1940s, as evidenced by Harvard photographic plate data. The collective properties of V4141~Sgr make it an intriguing target for continued monitoring in the coming weeks, months, and years.

\section{Observational data}

For this study, we have compiled photometric data of V4141~Sgr spanning the period from 1909 to 2025, albeit with significant gaps. We started monitoring the system in 2004, and regular survey observations are available from approximately 2014. The earliest segments of the light curve are covered by photographic plate observations from the Harvard College Observatory, accessible through the DASCH (Digital Access to a Sky Century at Harvard) archive \citep{2010AJ....140.1062L}. These data have been calibrated to APASS $B$-band magnitudes. Additionally, we incorporate the $V$-band observations reported by \citet{1989SvAL...15...13K,2001A&A...376..978K} and \citet{2025Galax..13....5K}\footnote{We noticed that some reported values from earlier epochs (until 1990) differ between the earlier \citep{2001A&A...376..978K} and later works \citep{2025Galax..13....5K}, and some are missing in the most recent paper. We have adopted the values from \citet{2001A&A...376..978K} for data until 2001 and from \citep{2025Galax..13....5K} for later epochs. However, this issue does not impact the discussion of the results in this work.}, and two $B$-band magnitudes from the USNO-A2.0 Catalogue \citep{1998yCat.1252....0M}.

Our data were obtained in $V$ and $I$ filters using 30-cm (until 2011) and 35-cm Schmidt–Cassegrain and Ritchey-Chretien telescopes equipped with SBIG CCD cameras at the Kleinkaroo Observatory in South Africa. The recent portion of the light curve is also covered by \textit{Gaia} DR3 epoch photometry \citep[$BP, G, RP$ filters;][]{2023A&A...674A...1G}, data from the All-Sky Automated Survey for Supernovae \citep[ASAS-SN;][]{2014ApJ...788...48S,2017PASP..129j4502K} in the $V$ and $g$ bands, and $o$- and $c$-band observations from the Asteroid Terrestrial-impact Last Alert System (ATLAS) project \citep{2018PASP..130f4505T,2020PASP..132h5002S}, obtained via the ATLAS Forced Photometry server \citep{2021TNSAN...7....1S}.

Following the detection of the recent outburst, we obtained three low-resolution optical spectra ($\sim$3600 -- 7800\AA) of V4141~Sgr using a remotely controlled 30-cm f/4 Newtonian telescope equipped with an Alpy600 spectrograph (R = 600), located at Deep Sky Chile near Cerro Tololo Observatory. The first spectrum, taken on February 7, 2025, was limited to a total exposure time of 2\,$\times$\,1200 s due to the low altitude of the star. The second and third spectra, acquired on March 1, 2025 and March 17, 2025, consist of a sum of five and six individual frames, each with an exposure time of 1200 s, respectively.

In addition to our new spectroscopic observations, we analyze the \textit{Gaia} low-resolution BP/RP slitless spectrum \citep[R $\sim$ 30–100; see][]{2023A&A...674A...2D,2023A&A...674A...3M}, published as part of \textit{Gaia} DR3, and obtained using {\tt GaiaXPy} package\footnote{\hyperlink{https://gaia-dpci.github.io/GaiaXPy-website/}{https://gaia-dpci.github.io/GaiaXPy-website/}}. Only the mean spectrum is available, derived from data collected between 2014 and 2017.

\begin{figure}[]
\centering
\includegraphics[width=\columnwidth]{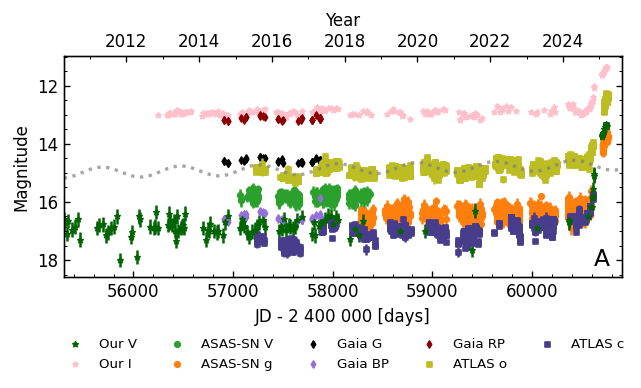}
\includegraphics[width=0.47\columnwidth]{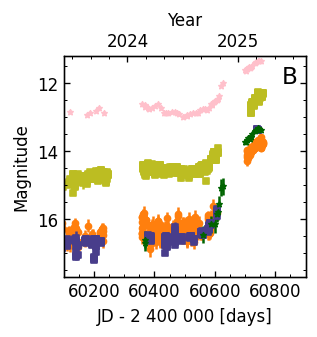}
\includegraphics[width=0.50\columnwidth]{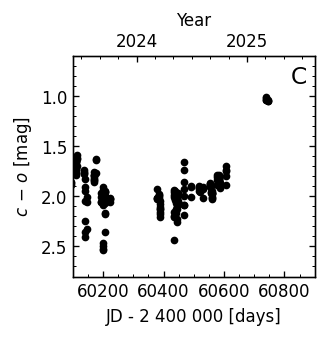}
\caption{Recent light curves of V4141~Sgr. \textbf{A:} Our photometric observations, together with data from ASAS-SN, ATLAS, and \textit{Gaia}, covering the period from $\sim$2010 to 2025. The dashed sinusoidal curve represents the apparent long-term variability of the star. \textbf{B:} Zoomed-in view of the photometric evolution during the recent brightening. \textbf{C:} Color evolution derived from ATLAS $c$- and $o$-band observations.}
\label{fig:recent_lc}
\end{figure}

\begin{figure}
\centering
\includegraphics[width=\columnwidth]{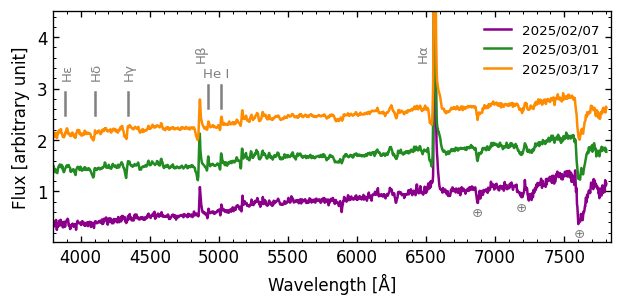}
\includegraphics[width=0.49\columnwidth]{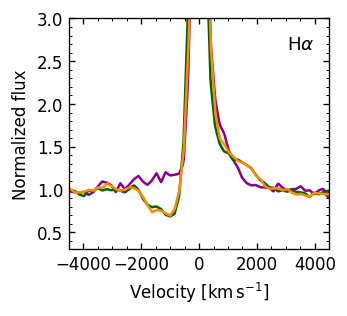}
\includegraphics[width=0.49\columnwidth]{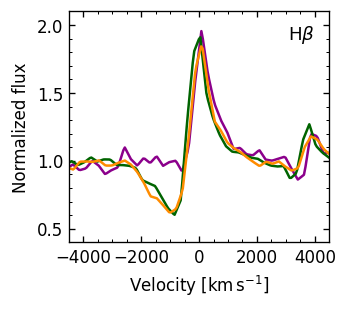}
\caption{Optical spectra of V4141~Sgr. The positions of strong emission lines are marked, with most of the unlabeled emission features attributed to \ion{Fe}{ii}. Telluric absorption features are indicated by the $\oplus$ symbol. The lower panels show the velocity profiles of H$\alpha$ and H$\beta$.}
\label{fig:spectra}
\end{figure}

\section{Results}

\subsection{Recent outburst activity}

We first noticed the brightening of V4141~Sgr in early February 2025\footnote{\hyperlink{http://www.cbat.eps.harvard.edu/unconf/followups/J17502384-1953422.html}{http://www.cbat.eps.harvard.edu/unconf/followups/J17502384-1953422.html}}, when the star became observable again after the seasonal gap. However, reanalysis of our photometric observations and data from ASAS-SN and ATLAS indicate that the outburst began as early as September 2024 (see Fig.~\ref{fig:recent_lc}A, B). Recent observations (our last data are from March 24, 2025) suggest that the brightness has not yet reached its peak.

As of mid-March 2025, the star has brightened by approximately 2.6--2.7 mag in the ATLAS $o$ and ASAS-SN $g$ bands and by 3.5 mag in the bluer ATLAS $c$ band compared to its average magnitude in quiescence (Fig.~\ref{fig:recent_lc}B). This wavelength-dependent amplitude suggests that the outburst is stronger at shorter wavelengths. Given the available photometric data, the only reliable color index that can be constructed is $c-o$, although post-seasonal gap ATLAS $c$-band observations remain sparse (Fig.~\ref{fig:recent_lc}C). The data confirms that the star appears significantly bluer during the outburst ($c-o \sim 1.0$ mag) compared to its long-term quiescent color ($c-o \sim 2.1$ mag).

Our spectroscopic observations, taken in February and March 2025, capture the star in its bright state. According to ASAS-SN $g$-band data, the brightness increased by approximately 0.3 mag between the first and second epoch and only by $\sim$0.05 mag until the third observation. Our first spectrum shows strong emission from Balmer lines, \ion{He}{i}, and \ion{Fe}{ii}, but no high-excitation emission features (Fig.~\ref{fig:spectra}) reported in the qiescent spectra \citep[e.g., {[\ion{O}{iii}]} or \ion{He}{ii};][]{2001A&A...376..978K}. The appearance of the star changed dramatically in the second observation, with the Balmer lines developing prominent P Cygni profiles, with broad blue-shifted absorption at velocities of approximately –1000 to –1500 km\,s$^{-1}$ (see lower panels of Fig.~\ref{fig:spectra}). Additionally, at least the H$\alpha$ line exhibits an extra redshifted emission component. The spectrum obtained in mid-March does not show prominent changes in comparison with the previous epoch.

\subsection{Historical photometry}

\begin{figure}
\centering
\includegraphics[width=\columnwidth]{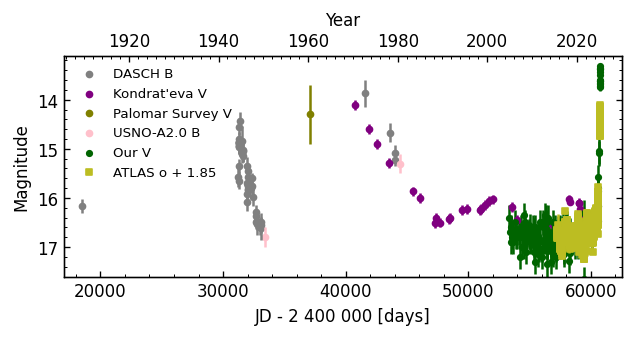}
\caption{Long-term photometry of V4141~Sgr. Older data include DASCH observations (calibrated to the $B$ band), $V$ magnitudes from \citet{2001A&A...376..978K}, including their assessment of the Palomar Survey value and from \citet{2025Galax..13....5K}, and $B$ magnitudes from the USNO-A2.0 catalog. The recent evolution is shown using our $V$ data and ATLAS $o$-band observations, which have been shifted by +1.85 mag for clarity. For better readability, other bands from recent epochs have been omitted.}
\label{fig:long_term}
\end{figure}

\begin{figure}
\centering
\includegraphics[width=\columnwidth]{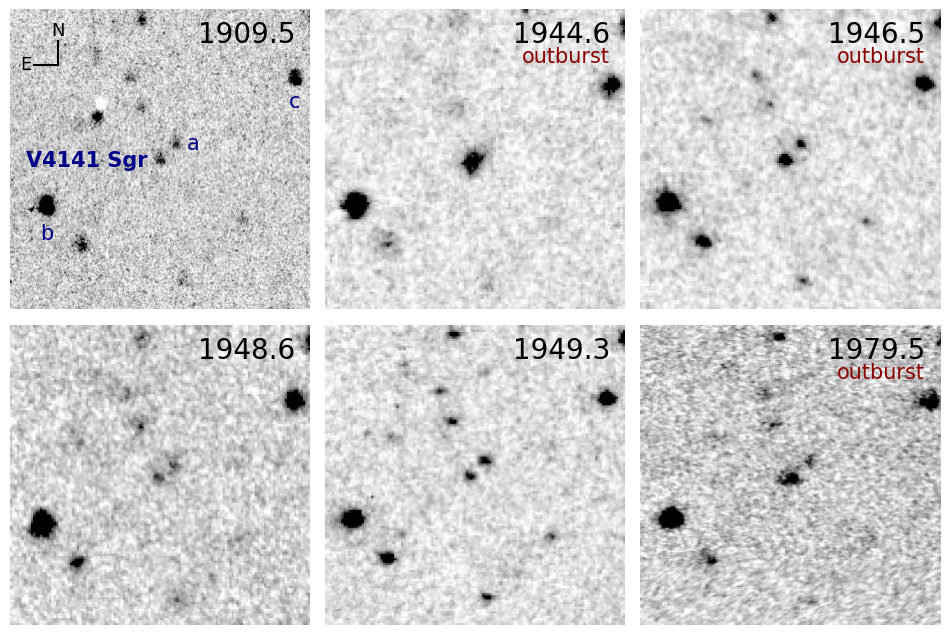}
\caption{DASCH photographic plates of V4141~Sgr at different epochs. The field of view is 120\arcsec $\times$ 120\arcsec. The position of V4141~Sgr is marked in the first image, along with three additional stars discussed in the text: (a) 2MASS J17502296-1953309, (b) 2MASS J17503040-1954169, and (c) 2MASS J17501606-1952414.}
\label{fig:dasch}
\end{figure}

The long-term variability of V4141~Sgr is illustrated in Fig.~\ref{fig:long_term}. The combined data from DASCH, \citet{2001A&A...376..978K,2025Galax..13....5K}, and recent surveys trace over a century of the photometric history of the system, albeit with significant gaps. It is important to note that the data in Fig.~\ref{fig:long_term} come from different filters (primarily $B$ and $V$ bands), and no magnitude shifts have been applied, apart from an arbitrary shift of ATLAS $o$ data. Therefore, while absolute values cannot be directly compared, the overall trends are clearly visible.

In addition to the recent outburst discussed in the previous section and the recovery from a brightening event between the 1960s–1970s and 1990, as reported by \citet{1989SvAL...15...13K,2001A&A...376..978K}, we identified an earlier brightening event captured on DASCH plates. Due to the sparsity of data, it is not possible to precisely determine when this outburst began or when the system returned to its quiescent state. However, the system was undoubtedly in a bright state during the photographic observations in 1944 and exhibited a slow decline in brightness at least until 1948 or 1949 (see Fig. \ref{fig:long_term}).

Figure \ref{fig:dasch} compares DASCH images of the V4141~Sgr field from 1909, during and after the 1940s brightening, and during the decline from the 1970s outburst. These images confirm that the 1940s brightening was a real outburst rather than an observational artifact. This is evidenced by the relative brightness of V4141~Sgr compared to its nearest bright neighbor (2MASS J17502296-1953309) and two additional bright stars in the field (2MASS J17503040-1954169 and 2MASS J17501606-1952414). On the photographic plate from 1946, V4141~Sgr appears significantly brighter than its northeast neighbor, similar to the 1979 observation.

The available data do not allow us to determine precisely when the 1970s outburst began. However, \citet{2001A&A...376..978K} reported that the system was already bright in 1960, based on observations from the Palomar Survey. This implies that the preceding quiescent phase lasted no more than approximately 11--12 years. Only mild increases in brightness were observed by \citet{2025Galax..13....5K} after the system reached its minimum luminosity in 1990. While their light curve extends up to 2024, it contains significant gaps, including between 2001 and 2005, as well as 2006 and 2017, with only a single observation in 2014. Our recent data, beginning in 2004, complement the later part of the light curve well, leaving only the period between 2001 and 2004 uncovered. However, V4141~Sgr is absent from the All Sky Automated Survey \citep[ASAS;][]{1997AcA....47..467P} catalog, which includes observations from 2000 to 2009. This suggests that no significant outburst occurred between 1990 and the recent event in 2024/2025, indicating that the system remained in quiescence for approximately 34--35 years.

\subsection{Quiescent spectral appearance and variability}
Although V4141~Sgr remained in a quiescent state for approximately three decades, little information is available regarding its quiescent appearance and variability, and even its infrared type (S or D) is unclear. Most previously published observations were conducted during the later stages of the 1970s outburst. The spectra analyzed by \citet{1997A&A...327..191M} were obtained in late April 1984, covering a wavelength range of 4000--7400\,\AA{}. Similarly, \citet{1992PASP..104.1187G} observed the star in July 1987, with their spectra spanning 3600--7100 \AA{}. Neither dataset revealed strong absorption bands.

The exact date of the spectra obtained by \citet{1984PASA....5..369A} is unclear. However, they were taken using the Faint Object Red Spectrograph on the Anglo-Australian Telescope, which was completed in 1984, indicating that the observations must have been conducted either during the commissioning time in 1983 or early 1984. The instrument covered wavelengths from 5000 to 10\,000 \AA{}, which likely facilitated the detection of M-type absorption features.

The only spectra definitively obtained during quiescence are those presented by \citet{2001A&A...376..978K} and \citet{2025Galax..13....5K}, though these are low-resolution and not discussed in detail. Only some of the spectra taken after 1990 extend beyond 5100 \AA{} (up to 7300\AA{}) but do not show TiO bands. \citet{2005A&A...435.1087L} obtained a spectrum in June 2002, when the system was already in quiescence, though they did not discuss the continuum. However, they reported the presence of \ion{He}{ii} lines, with no indications of emission lines from species with even higher ionization potentials, such as [\ion{Fe}{vii}] or Raman-scattered \ion{O}{vi}.

\begin{figure}
\centering
\includegraphics[width=\columnwidth]{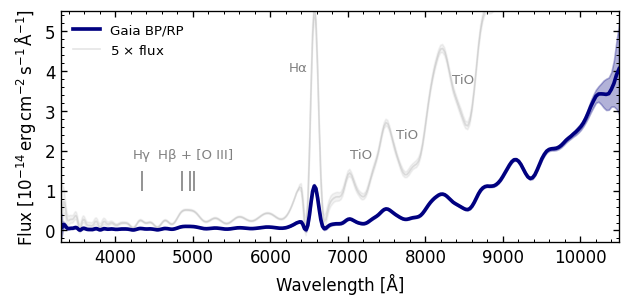}
\caption{\textit{Gaia} BP/RP spectrum of V4141~Sgr. The gray curve shows the flux multiplied by a factor of five to enhance the visibility of fainter features. Flux uncertainties, as reported in the \textit{Gaia} DR3 catalog, are shown as shaded regions. The most prominent emission and absorption lines/bands are marked.}
\label{fig:gaiaxp}
\end{figure}

A low-resolution \textit{Gaia} BP/RP spectrum of V4141~Sgr was published in \textit{Gaia} DR3 (Fig. \ref{fig:gaiaxp}). This spectrum represents the median flux over the \textit{Gaia} DR3 observing period from 2014 to 2017 when the star was in quiescence. It clearly exhibits the continuum of an M-type star, characterized by prominent TiO bands overlaid with strong emission lines, resembling the typical spectrum of an S-type symbiotic star. Although the resolution does not allow for detailed line identification, the presence of H$\alpha$, H$\beta$, H$\gamma$, and likely [\ion{O}{iii}] emission is evident.

Part of the quiescent phase preceding the 2024/2025 outburst is covered by our photometry and data from ATLAS, ASAS-SN, and \textit{Gaia}. Due to the relatively low brightness of the star during quiescence, the data exhibit significant scatter, particularly in the case of ASAS-SN and our $V$ data. However, both ATLAS and \textit{Gaia} DR3 data (though the latter spans only $\sim$1000 days) reveal a low-amplitude, seemingly sinusoidal variability with a period of approximately 790 days. This variability could be due to the orbital motion of V4141~Sgr, as similar timescales are typical for the orbital periods of S-type symbiotic stars \citep[e.g.,][]{2012BaltA..21....5M,2013AcA....63..405G,2019RNAAS...3...28M}. However, only spectroscopic follow-up can definitively confirm its nature. We also note that neither our photometry nor data from other surveys confirm any specific brightening event in 2018, as reported by \citet{2025Galax..13....5K}, beyond the observed long-term variability.

To conclude the discussion on the infrared classification, V4141~Sgr was observed by 2MASS during its quiescent phase in September 2000. Its infrared colors, $J-H$ = 1.2 mag and $H-K_{\rm s}$ = 0.6 mag, are consistent with those of S-type symbiotic stars. Moreover, the published [\ion{O}{iii}] observations, namely [\ion{O}{iii}]\,$\lambda$4363\,\AA{}/H$\gamma$ = 0.81 \citep[1987; ][]{1992PASP..104.1187G} and
0.64 \citep[2002; ][]{2005A&A...435.1087L}, and [\ion{O}{iii}]\,$\lambda$5007\,\AA{}/H$\beta$ = 1.27 (1987) and 0.96 (2002), locate V4141~Sgr among S-type symbiotic stars (and very far from PNe) in the [\ion{O}{iii}] diagnostic diagram \citep[see][]{1995PASP..107..462G,2017A&A...606A.110I}.

\section{Discussion}
\subsection{Detection of optically-thick outflows}

The detection of P Cygni profiles in optical spectra, indicative of high-velocity, optically-thick outflows, is not unexpected during symbiotic star outbursts. However, such observations remain relatively scarce, likely due to their inclination dependence and their appearance being restricted to specific outburst phases. A~similar spectral evolution to that observed in V4141 Sgr has been reported during the outbursts of several other symbiotic stars, including Z And \citep[e.g.,][]{2008MNRAS.389..829T}, BF Cyg \citep[e.g.,][]{2019CoSka..49..411S}, CH Cyg \citep[e.g.,][]{1996MNRAS.278..542T,2019A&A...622A..45I}, CI Cyg \citep[e.g.,][]{1975IAUS...67..401M}, AG Peg \citep{1942ApJ....95..386M,1993AJ....106.1573K}, RX Pup \citep[e.g.,][]{1976A&A....46..303S}, and Hen 3-1341 \citep{2000A&A...354L..25T}. P Cygni profiles are also commonly observed during the early outburst phases of symbiotic recurrent novae. A particularly prolonged presence of these profiles was reported in the peculiar symbiotic system V694 Mon \citep[e.g.,][]{2020MNRAS.492.3107L}.

Additionally, blue- and red-shifted satellite emission components, attributed to bipolar jets, have been detected in the optical spectra of some systems, though such detections remain rare. They were reported in optical spectra of Z And \citep[e.g.,][]{2007MNRAS.376L..16T,2007A&A...461L...5B,2009ApJ...690.1222S}, St~2-22 \citep{2017AcA....67..225T,2022A&A...657A.137G}, Hen 3-1341 \citep{2000A&A...354L..25T}, StH$\alpha$ 190 \citep{2001A&A...369L...1M}, and BF Cyg \citep{2013A&A...551L..10S}.

Overall, our observations place V4141 Sgr among the small group of symbiotic stars where optical spectra reveal evidence of outflows and, likely, bipolar jets.

\subsection{Recurring outbursts of V4141~Sgr}
Our detection of the ongoing outburst of V4141~Sgr, along with the historical outburst we uncovered on Harvard photographic plates and the brightening reported by \citet{1989SvAL...15...13K,2001A&A...376..978K}, makes this system particularly intriguing. Notably, the spectroscopic evolution of V4141~Sgr during its 1970s outburst, as well as its timescale, resembles the behavior of "slow" symbiotic novae such as RR Tel, AG Peg, and PU Vul.

This is not the first instance where a later eruption followed a "slow" symbiotic nova outburst. The slowest known symbiotic nova, AG Peg, underwent a Z And-type outburst in 2015, 165 years after its nova brightening \citep{2016MNRAS.462.4435T,2017A&A...604A..48S,2019CoSka..49..228M}. More recently, a similar pattern was observed in V426 Sge, which experienced a symbiotic nova outburst in 1968 followed by a Z And-type outburst in 2018 \citep{2020A&A...636A..77S}. However, no prior outbursts were reported for these systems before their nova events. In the case of V426 Sge, no signs of symbiotic activity were observed since the early 20th century until its outburst \citep{2020A&A...636A..77S}.

On the other hand, \citet{2023MNRAS.523..163M} reported that V618 Sgr is currently undergoing an outburst with an evolution characteristic of "slow" symbiotic novae, and their analysis of historical data revealed additional brightenings in the past. However, their nature remains uncertain due to the lack of spectroscopic data. RX Pup may have also experienced an outburst around 1894 before its nova outburst in the 1970s, as suggested by \cite{1999MNRAS.305..190M}, though this remains inconclusive. Finally, LMC S154 appears to have undergone multiple symbiotic nova outbursts, leading \citet{2019A&A...624A.133I} to classify it as a symbiotic recurrent nova. However, the duration of its outbursts differs from those of other symbiotic recurrent novae, such as T CrB, RS Oph, and V3890 Sgr. 

In any case, the photometric history of V4141~Sgr makes it unique among symbiotic stars, and its current outburst is undoubtedly worth follow-up observations. This is particularly compelling given that the ongoing brightening is accompanied by a significant increase in the luminosity of the hot component (see Appendix \ref{app:hot}), whereas typical outbursts of classical symbiotic stars—such as Z And, CI Cyg, or AX Per—generally exhibit a nearly constant luminosity throughout the outburst \citep[e.g., Fig. 1 in][]{2010arXiv1011.5657M}.

\section{Conclusions}

Our observations of V4141~Sgr during its recent outburst reveal the presence of high-velocity, optically thick outflows, with possible evidence for bipolar jets. This places the system among the small group of symbiotic stars where such features have been detected in optical spectra. The historical light curve further indicates that V4141~Sgr has experienced at least two additional outbursts over the past century. The current outburst, following approximately 35 years of quiescence, is preceded by a prolonged outburst in the 1970s, whose spectroscopic evolution resembled that of "slow" symbiotic novae. These recurring outbursts and the detection of P Cygni profiles make V4141~Sgr a particularly intriguing system, warranting continued photometric and spectroscopic monitoring in the future.

\begin{acknowledgements}
We thank the anonymous referee for their helpful comments that improved the manuscript. The research of JaM was supported by the Czech Science Foundation (GACR) project no. 24-10608O and by the Spanish Ministry of Science and Innovation with the grant no. PID2023-146453NB-100 (PLAtoSOnG). JMik was supported by the Polish National Science Centre (NCN) grant 2023/48/Q/ST9/00138.
\end{acknowledgements}

%
  \bibliographystyle{aa} 
  \bibliography{bibliography} 

\begin{thebibliography}{60}
\expandafter\ifx\csname natexlab\endcsname\relax\def\natexlab#1{#1}\fi

\bibitem[{{Akras} {et~al.}(2019){Akras}, {Guzman-Ramirez}, {Leal-Ferreira}, \& {Ramos-Larios}}]{2019ApJS..240...21A}
{Akras}, S., {Guzman-Ramirez}, L., {Leal-Ferreira}, M.~L., \& {Ramos-Larios}, G. 2019, \apjs, 240, 21

\bibitem[{{Allen}(1984)}]{1984PASA....5..369A}
{Allen}, D.~A. 1984, \pasa, 5, 369

\bibitem[{{Bailer-Jones} {et~al.}(2021){Bailer-Jones}, {Rybizki}, {Fouesneau}, {Demleitner}, \& {Andrae}}]{2021AJ....161..147B}
{Bailer-Jones}, C.~A.~L., {Rybizki}, J., {Fouesneau}, M., {Demleitner}, M., \& {Andrae}, R. 2021, \aj, 161, 147

\bibitem[{{Belczy{\'n}ski} {et~al.}(2000){Belczy{\'n}ski}, {Miko{\l}ajewska}, {Munari}, {Ivison}, \& {Friedjung}}]{2000A&AS..146..407B}
{Belczy{\'n}ski}, K., {Miko{\l}ajewska}, J., {Munari}, U., {Ivison}, R.~J., \& {Friedjung}, M. 2000, \aaps, 146, 407

\bibitem[{{Burmeister} \& {Leedj{\"a}rv}(2007)}]{2007A&A...461L...5B}
{Burmeister}, M. \& {Leedj{\"a}rv}, L. 2007, \aap, 461, L5

\bibitem[{{De Angeli} {et~al.}(2023){De Angeli}, {Weiler}, {Montegriffo}, {Evans}, {Riello}, {Andrae}, {Carrasco}, {Busso}, {Burgess}, {Cacciari}, {Davidson}, {Harrison}, {Hodgkin}, {Jordi}, {Osborne}, {Pancino}, {Altavilla}, {Barstow}, {Bailer-Jones}, {Bellazzini}, {Brown}, {Castellani}, {Cowell}, {Delchambre}, {De Luise}, {Diener}, {Fabricius}, {Fouesneau}, {Fr{\'e}mat}, {Gilmore}, {Giuffrida}, {Hambly}, {Hidalgo}, {Holland}, {Kostrzewa-Rutkowska}, {van Leeuwen}, {Lobel}, {Marinoni}, {Miller}, {Pagani}, {Palaversa}, {Piersimoni}, {Pulone}, {Ragaini}, {Rainer}, {Richards}, {Rixon}, {Ruz-Mieres}, {Sanna}, {Sarro}, {Rowell}, {Sordo}, {Walton}, \& {Yoldas}}]{2023A&A...674A...2D}
{De Angeli}, F., {Weiler}, M., {Montegriffo}, P., {et~al.} 2023, \aap, 674, A2

\bibitem[{{Gaia Collaboration} {et~al.}(2023){Gaia Collaboration}, {Vallenari}, {Brown}, {Prusti}, {de Bruijne}, {Arenou}, {Babusiaux}, {Biermann}, {Creevey}, {Ducourant}, {Evans}, {Eyer}, {Guerra}, {Hutton}, {Jordi}, {Klioner}, {Lammers}, {Lindegren}, {Luri}, {Mignard}, {Panem}, {Pourbaix}, {Randich}, {Sartoretti}, {Soubiran}, {Tanga}, {Walton}, {Bailer-Jones}, {Bastian}, {Drimmel}, {Jansen}, {Katz}, {Lattanzi}, {van Leeuwen}, {Bakker}, {Cacciari}, {Casta{\~n}eda}, {De Angeli}, {Fabricius}, {Fouesneau}, {Fr{\'e}mat}, {Galluccio}, {Guerrier}, {Heiter}, {Masana}, {Messineo}, {Mowlavi}, {Nicolas}, {Nienartowicz}, {Pailler}, {Panuzzo}, {Riclet}, {Roux}, {Seabroke}, {Sordo}, {Th{\'e}venin}, {Gracia-Abril}, {Portell}, {Teyssier}, {Altmann}, {Andrae}, {Audard}, {Bellas-Velidis}, {Benson}, {Berthier}, {Blomme}, {Burgess}, {Busonero}, {Busso}, {C{\'a}novas}, {Carry}, {Cellino}, {Cheek}, {Clementini}, {Damerdji}, {Davidson}, {de Teodoro}, {Nu{\~n}ez Campos}, {Delchambre}, {Dell'Oro}, {Esquej},
  {Fern{\'a}ndez-Hern{\'a}ndez}, {Fraile}, {Garabato}, {Garc{\'\i}a-Lario}, {Gosset}, {Haigron}, {Halbwachs}, {Hambly}, {Harrison}, {Hern{\'a}ndez}, {Hestroffer}, {Hodgkin}, {Holl}, {Jan{\ss}en}, {Jevardat de Fombelle}, {Jordan}, {Krone-Martins}, {Lanzafame}, {L{\"o}ffler}, {Marchal}, {Marrese}, {Moitinho}, {Muinonen}, {Osborne}, {Pancino}, {Pauwels}, {Recio-Blanco}, {Reyl{\'e}}, {Riello}, {Rimoldini}, {Roegiers}, {Rybizki}, {Sarro}, {Siopis}, {Smith}, {Sozzetti}, {Utrilla}, {van Leeuwen}, {Abbas}, {{\'A}brah{\'a}m}, {Abreu Aramburu}, {Aerts}, {Aguado}, {Ajaj}, {Aldea-Montero}, {Altavilla}, {{\'A}lvarez}, {Alves}, {Anders}, {Anderson}, {Anglada Varela}, {Antoja}, {Baines}, {Baker}, {Balaguer-N{\'u}{\~n}ez}, {Balbinot}, {Balog}, {Barache}, {Barbato}, {Barros}, {Barstow}, {Bartolom{\'e}}, {Bassilana}, {Bauchet}, {Becciani}, {Bellazzini}, {Berihuete}, {Bernet}, {Bertone}, {Bianchi}, {Binnenfeld}, {Blanco-Cuaresma}, {Blazere}, {Boch}, {Bombrun}, {Bossini}, {Bouquillon}, {Bragaglia}, {Bramante}, {Breedt},
  {Bressan}, {Brouillet}, {Brugaletta}, {Bucciarelli}, {Burlacu}, {Butkevich}, {Buzzi}, {Caffau}, {Cancelliere}, {Cantat-Gaudin}, {Carballo}, {Carlucci}, {Carnerero}, {Carrasco}, {Casamiquela}, {Castellani}, {Castro-Ginard}, {Chaoul}, {Charlot}, {Chemin}, {Chiaramida}, {Chiavassa}, {Chornay}, {Comoretto}, {Contursi}, {Cooper}, {Cornez}, {Cowell}, {Crifo}, {Cropper}, {Crosta}, {Crowley}, {Dafonte}, {Dapergolas}, {David}, {David}, {de Laverny}, {De Luise}, \& {De March}}]{2023A&A...674A...1G}
{Gaia Collaboration}, {Vallenari}, A., {Brown}, A.~G.~A., {et~al.} 2023, \aap, 674, A1

\bibitem[{{Ga{\l}an} {et~al.}(2022){Ga{\l}an}, {Miko{\l}ajewska}, {I{\l}kiewicz}, {Monard}, {{\.Z}ywica}, \& {Zamanov}}]{2022A&A...657A.137G}
{Ga{\l}an}, C., {Miko{\l}ajewska}, J., {I{\l}kiewicz}, K., {et~al.} 2022, \aap, 657, A137

\bibitem[{{Gromadzki} {et~al.}(2013){Gromadzki}, {Miko{\l}ajewska}, \& {Soszy{\'n}ski}}]{2013AcA....63..405G}
{Gromadzki}, M., {Miko{\l}ajewska}, J., \& {Soszy{\'n}ski}, I. 2013, \actaa, 63, 405

\bibitem[{{Gutierrez-Moreno} {et~al.}(1992){Gutierrez-Moreno}, {Moreno}, \& {Cortes}}]{1992PASP..104.1187G}
{Gutierrez-Moreno}, A., {Moreno}, H., \& {Cortes}, G. 1992, \pasp, 104, 1187

\bibitem[{{Gutierrez-Moreno} {et~al.}(1995){Gutierrez-Moreno}, {Moreno}, \& {Cortes}}]{1995PASP..107..462G}
{Gutierrez-Moreno}, A., {Moreno}, H., \& {Cortes}, G. 1995, \pasp, 107, 462

\bibitem[{{Iijima} {et~al.}(2019){Iijima}, {Naito}, \& {Narusawa}}]{2019A&A...622A..45I}
{Iijima}, T., {Naito}, H., \& {Narusawa}, S. 2019, \aap, 622, A45

\bibitem[{{I{\l}kiewicz} \& {Miko{\l}ajewska}(2017)}]{2017A&A...606A.110I}
{I{\l}kiewicz}, K. \& {Miko{\l}ajewska}, J. 2017, \aap, 606, A110

\bibitem[{{I{\l}kiewicz} {et~al.}(2019){I{\l}kiewicz}, {Miko{\l}ajewska}, {Miszalski}, {Gromadzki}, {Monard}, \& {Amigo}}]{2019A&A...624A.133I}
{I{\l}kiewicz}, K., {Miko{\l}ajewska}, J., {Miszalski}, B., {et~al.} 2019, \aap, 624, A133

\bibitem[{{Kenyon} {et~al.}(1993){Kenyon}, {Miko{\l}ajewska}, {Mikolajewski}, {Polidan}, \& {Slovak}}]{1993AJ....106.1573K}
{Kenyon}, S.~J., {Miko{\l}ajewska}, J., {Mikolajewski}, M., {Polidan}, R.~S., \& {Slovak}, M.~H. 1993, \aj, 106, 1573

\bibitem[{{Kochanek} {et~al.}(2017){Kochanek}, {Shappee}, {Stanek}, {Holoien}, {Thompson}, {Prieto}, {Dong}, {Shields}, {Will}, {Britt}, {Perzanowski}, \& {Pojma{\'n}ski}}]{2017PASP..129j4502K}
{Kochanek}, C.~S., {Shappee}, B.~J., {Stanek}, K.~Z., {et~al.} 2017, \pasp, 129, 104502

\bibitem[{{Kondrat'eva} {et~al.}(2025){Kondrat'eva}, {Denissyuk}, {Shomshekova}, {Reva}, {Aimanova}, \& {Krugov}}]{2025Galax..13....5K}
{Kondrat'eva}, L., {Denissyuk}, E., {Shomshekova}, S., {et~al.} 2025, Galaxies, 13, 5

\bibitem[{{Kondrat'eva}(1975)}]{1975TrAlm..25...23K}
{Kondrat'eva}, L.~N. 1975, Trudy Astrofizicheskogo Instituta Alma-Ata, 25, 23

\bibitem[{{Kondrat'eva}(1989)}]{1989SvAL...15...13K}
{Kondrat'eva}, L.~N. 1989, Soviet Astronomy Letters, 15, 13

\bibitem[{{Kondrat'eva}(2001)}]{2001A&A...376..978K}
{Kondrat'eva}, L.~N. 2001, \aap, 376, 978

\bibitem[{{Laycock} {et~al.}(2010){Laycock}, {Tang}, {Grindlay}, {Los}, {Simcoe}, \& {Mink}}]{2010AJ....140.1062L}
{Laycock}, S., {Tang}, S., {Grindlay}, J., {et~al.} 2010, \aj, 140, 1062

\bibitem[{{Lucy} {et~al.}(2020){Lucy}, {Sokoloski}, {Munari}, {Roy}, {Kuin}, {Rupen}, {Knigge}, {Darnley}, {Luna}, {Somogyi}, {Valisa}, {Milani}, {Sollecchia}, \& {Weston}}]{2020MNRAS.492.3107L}
{Lucy}, A.~B., {Sokoloski}, J.~L., {Munari}, U., {et~al.} 2020, \mnras, 492, 3107

\bibitem[{{Luna} \& {Costa}(2005)}]{2005A&A...435.1087L}
{Luna}, G.~J.~M. \& {Costa}, R.~D.~D. 2005, \aap, 435, 1087

\bibitem[{{MacConnell}(1983)}]{1983RMxAA...8...39M}
{MacConnell}, D.~J. 1983, \rmxaa, 8, 39

\bibitem[{{Mammano} {et~al.}(1975){Mammano}, {Rosino}, \& {Yildizdogdu}}]{1975IAUS...67..401M}
{Mammano}, A., {Rosino}, L., \& {Yildizdogdu}, S. 1975, in IAU Symposium, Vol.~67, Variable Stars and Stellar Evolution, ed. V.~E. {Sherwood} \& L.~{Plaut}, 401

\bibitem[{{Merc} {et~al.}(2019{\natexlab{a}}){Merc}, {G{\'a}lis}, \& {Teyssier}}]{2019CoSka..49..228M}
{Merc}, J., {G{\'a}lis}, R., \& {Teyssier}, F. 2019{\natexlab{a}}, Contributions of the Astronomical Observatory Skalnate Pleso, 49, 228

\bibitem[{{Merc} {et~al.}(2023){Merc}, {G{\'a}lis}, {Velez}, {Charbonnel}, {Garde}, {Le D{\^u}}, {Mulato}, {Petit}, {Bohlsen}, {Curry}, {Love}, \& {Barker}}]{2023MNRAS.523..163M}
{Merc}, J., {G{\'a}lis}, R., {Velez}, P., {et~al.} 2023, \mnras, 523, 163

\bibitem[{{Merc} {et~al.}(2019{\natexlab{b}}){Merc}, {G{\'a}lis}, \& {Wolf}}]{2019RNAAS...3...28M}
{Merc}, J., {G{\'a}lis}, R., \& {Wolf}, M. 2019{\natexlab{b}}, Research Notes of the American Astronomical Society, 3, 28

\bibitem[{{Merc} {et~al.}(2019{\natexlab{c}}){Merc}, {G{\'a}lis}, \& {Wolf}}]{2019AN....340..598M}
{Merc}, J., {G{\'a}lis}, R., \& {Wolf}, M. 2019{\natexlab{c}}, Astronomische Nachrichten, 340, 598

\bibitem[{{Merc} {et~al.}(2020){Merc}, {Miko{\l}ajewska}, {Gromadzki}, {Ga{\l}an}, {I{\l}kiewicz}, {Skowron}, {Wyrzykowski}, {Hodgkin}, {Rybicki}, {Zieli{\'n}ski}, {Kruszy{\'n}ska}, {Godunova}, {Simon}, {Reshetnyk}, {Lewis}, {Kolb}, {Morrell}, {Norton}, {Awiphan}, {Poshyachinda}, {Reichart}, {Greet}, \& {Kolgjini}}]{2020A&A...644A..49M}
{Merc}, J., {Miko{\l}ajewska}, J., {Gromadzki}, M., {et~al.} 2020, \aap, 644, A49

\bibitem[{{Merrill}(1942)}]{1942ApJ....95..386M}
{Merrill}, P.~W. 1942, \apj, 95, 386

\bibitem[{{Miko{\l}ajewska}(2010)}]{2010arXiv1011.5657M}
{Miko{\l}ajewska}, J. 2010, arXiv e-prints, arXiv:1011.5657

\bibitem[{{Miko{\l}ajewska}(2012)}]{2012BaltA..21....5M}
{Miko{\l}ajewska}, J. 2012, Baltic Astronomy, 21, 5

\bibitem[{{Miko{\l}ajewska} {et~al.}(1997){Miko{\l}ajewska}, {Acker}, \& {Stenholm}}]{1997A&A...327..191M}
{Miko{\l}ajewska}, J., {Acker}, A., \& {Stenholm}, B. 1997, \aap, 327, 191

\bibitem[{{Miko{\l}ajewska} {et~al.}(1999){Miko{\l}ajewska}, {Brandi}, {Hack}, {Whitelock}, {Barba}, {Garcia}, \& {Marang}}]{1999MNRAS.305..190M}
{Miko{\l}ajewska}, J., {Brandi}, E., {Hack}, W., {et~al.} 1999, \mnras, 305, 190

\bibitem[{{Miko{\l}ajewska} {et~al.}(1995){Miko{\l}ajewska}, {Kenyon}, {Mikolajewski}, {Garcia}, \& {Polidan}}]{1995AJ....109.1289M}
{Miko{\l}ajewska}, J., {Kenyon}, S.~J., {Mikolajewski}, M., {Garcia}, M.~R., \& {Polidan}, R.~S. 1995, \aj, 109, 1289

\bibitem[{{Monet} {et~al.}(1998){Monet}, {Bird}, {Canzian}, {Dahn}, {Guetter}, {Harris}, {Henden}, {Levine}, {Luginbuhl}, {Monet}, {Rhodes}, {Riepe}, {Sell}, {Stone}, {Vrba}, \& {Walker}}]{1998yCat.1252....0M}
{Monet}, A., {Bird}, J., {Canzian}, B., {et~al.} 1998, {VizieR Online Data Catalog: The USNO-A2.0 Catalogue}, I/252

\bibitem[{{Montegriffo} {et~al.}(2023){Montegriffo}, {De Angeli}, {Andrae}, {Riello}, {Pancino}, {Sanna}, {Bellazzini}, {Evans}, {Carrasco}, {Sordo}, {Busso}, {Cacciari}, {Jordi}, {van Leeuwen}, {Vallenari}, {Altavilla}, {Barstow}, {Brown}, {Burgess}, {Castellani}, {Cowell}, {Davidson}, {De Luise}, {Delchambre}, {Diener}, {Fabricius}, {Fr{\'e}mat}, {Fouesneau}, {Gilmore}, {Giuffrida}, {Hambly}, {Harrison}, {Hidalgo}, {Hodgkin}, {Holland}, {Marinoni}, {Osborne}, {Pagani}, {Palaversa}, {Piersimoni}, {Pulone}, {Ragaini}, {Rainer}, {Richards}, {Rowell}, {Ruz-Mieres}, {Sarro}, {Walton}, \& {Yoldas}}]{2023A&A...674A...3M}
{Montegriffo}, P., {De Angeli}, F., {Andrae}, R., {et~al.} 2023, \aap, 674, A3

\bibitem[{{Munari} {et~al.}(2001){Munari}, {Tomov}, {Yudin}, {Marrese}, {Zwitter}, {Gratton}, {Bonanno}, {Bruno}, {Cal{\'\i}}, {Claudi}, {Cosentino}, {Desidera}, {Farisato}, {Martorana}, {Marino}, {Rebeschini}, {Scuderi}, \& {Timpanaro}}]{2001A&A...369L...1M}
{Munari}, U., {Tomov}, T., {Yudin}, B.~F., {et~al.} 2001, \aap, 369, L1

\bibitem[{{Perek} \& {Kohoutek}(1967)}]{1967cgpn.book.....P}
{Perek}, L. \& {Kohoutek}, L. 1967, {Catalogue of Galactic Planetary Nebulae}

\bibitem[{{Pojmanski}(1997)}]{1997AcA....47..467P}
{Pojmanski}, G. 1997, \actaa, 47, 467

\bibitem[{{Schmidt} \& {Miko{\l}ajewska}(2003)}]{2003ASPC..303..163S}
{Schmidt}, M.~R. \& {Miko{\l}ajewska}, J. 2003, in Astronomical Society of the Pacific Conference Series, Vol. 303, Symbiotic Stars Probing Stellar Evolution, ed. R.~L.~M. {Corradi}, J.~{Miko{\l}ajewska}, \& T.~J. {Mahoney}, 163

\bibitem[{{Shappee} {et~al.}(2014){Shappee}, {Prieto}, {Grupe}, {Kochanek}, {Stanek}, {De Rosa}, {Mathur}, {Zu}, {Peterson}, {Pogge}, {Komossa}, {Im}, {Jencson}, {Holoien}, {Basu}, {Beacom}, {Szczygie{\l}}, {Brimacombe}, {Adams}, {Campillay}, {Choi}, {Contreras}, {Dietrich}, {Dubberley}, {Elphick}, {Foale}, {Giustini}, {Gonzalez}, {Hawkins}, {Howell}, {Hsiao}, {Koss}, {Leighly}, {Morrell}, {Mudd}, {Mullins}, {Nugent}, {Parrent}, {Phillips}, {Pojmanski}, {Rosing}, {Ross}, {Sand}, {Terndrup}, {Valenti}, {Walker}, \& {Yoon}}]{2014ApJ...788...48S}
{Shappee}, B.~J., {Prieto}, J.~L., {Grupe}, D., {et~al.} 2014, \apj, 788, 48

\bibitem[{{Shchurova} {et~al.}(2019){Shchurova}, {Skopal}, {Shugarov}, {Seker{\'a}{\v{s}}}, {Kom{\v{z}}{\'\i}k}, {Kundra}, \& {Shagatova}}]{2019CoSka..49..411S}
{Shchurova}, A., {Skopal}, A., {Shugarov}, S.~Y., {et~al.} 2019, Contributions of the Astronomical Observatory Skalnate Pleso, 49, 411

\bibitem[{{Shingles} {et~al.}(2021){Shingles}, {Smith}, {Young}, {Smartt}, {Tonry}, {Denneau}, {Heinze}, {Weiland}, {Flewelling}, {Stalder}, {Clocchiatti}, {F{\"o}rster}, {Pignata}, {Rest}, {Anderson}, {Stubbs}, \& {Erasmus}}]{2021TNSAN...7....1S}
{Shingles}, L., {Smith}, K.~W., {Young}, D.~R., {et~al.} 2021, Transient Name Server AstroNote, 7, 1

\bibitem[{{Skopal} {et~al.}(2009){Skopal}, {Pribulla}, {Budaj}, {Vittone}, {Errico}, {Wolf}, {Otsuka}, {Chrastina}, \& {Mikul{\'a}{\v{s}}ek}}]{2009ApJ...690.1222S}
{Skopal}, A., {Pribulla}, T., {Budaj}, J., {et~al.} 2009, \apj, 690, 1222

\bibitem[{{Skopal} {et~al.}(2020){Skopal}, {Shugarov}, {Munari}, {Masetti}, {Marchesini}, {Kom{\v{z}}{\'\i}k}, {Kundra}, {Shagatova}, {Tarasova}, {Buil}, {Boussin}, {Shenavrin}, {Hambsch}, {Dallaporta}, {Frigo}, {Garde}, {Zubareva}, {Dubovsk{\'y}}, \& {Kroll}}]{2020A&A...636A..77S}
{Skopal}, A., {Shugarov}, S.~Y., {Munari}, U., {et~al.} 2020, \aap, 636, A77

\bibitem[{{Skopal} {et~al.}(2017){Skopal}, {Shugarov}, {Seker{\'a}{\v{s}}}, {Wolf}, {Tarasova}, {Teyssier}, {Fujii}, {Guarro}, {Garde}, {Graham}, {Lester}, {Bouttard}, {Lemoult}, {Sollecchia}, {Montier}, \& {Boyd}}]{2017A&A...604A..48S}
{Skopal}, A., {Shugarov}, S.~Y., {Seker{\'a}{\v{s}}}, M., {et~al.} 2017, \aap, 604, A48

\bibitem[{{Skopal} {et~al.}(2013){Skopal}, {Tomov}, \& {Tomova}}]{2013A&A...551L..10S}
{Skopal}, A., {Tomov}, N.~A., \& {Tomova}, M.~T. 2013, \aap, 551, L10

\bibitem[{{Smith} {et~al.}(2020){Smith}, {Smartt}, {Young}, {Tonry}, {Denneau}, {Flewelling}, {Heinze}, {Weiland}, {Stalder}, {Rest}, {Stubbs}, {Anderson}, {Chen}, {Clark}, {Do}, {F{\"o}rster}, {Fulton}, {Gillanders}, {McBrien}, {O'Neill}, {Srivastav}, \& {Wright}}]{2020PASP..132h5002S}
{Smith}, K.~W., {Smartt}, S.~J., {Young}, D.~R., {et~al.} 2020, \pasp, 132, 085002

\bibitem[{{Stenholm} \& {Acker}(1987)}]{1987A&AS...68...51S}
{Stenholm}, B. \& {Acker}, A. 1987, \aaps, 68, 51

\bibitem[{{Swings} \& {Klutz}(1976)}]{1976A&A....46..303S}
{Swings}, J.~P. \& {Klutz}, M. 1976, \aap, 46, 303

\bibitem[{{The}(1964)}]{1964CoBos..28....1T}
{The}, P.~S. 1964, Contributions from the Bosscha Observervatory, 28, 1

\bibitem[{{Tomov} {et~al.}(2007){Tomov}, {Tomova}, \& {Bisikalo}}]{2007MNRAS.376L..16T}
{Tomov}, N.~A., {Tomova}, M.~T., \& {Bisikalo}, D.~V. 2007, \mnras, 376, L16

\bibitem[{{Tomov} {et~al.}(2008){Tomov}, {Tomova}, \& {Bisikalo}}]{2008MNRAS.389..829T}
{Tomov}, N.~A., {Tomova}, M.~T., \& {Bisikalo}, D.~V. 2008, \mnras, 389, 829

\bibitem[{{Tomov} {et~al.}(1996){Tomov}, {Kolev}, {Munari}, \& {Antov}}]{1996MNRAS.278..542T}
{Tomov}, T., {Kolev}, D., {Munari}, U., \& {Antov}, A. 1996, \mnras, 278, 542

\bibitem[{{Tomov} {et~al.}(2000){Tomov}, {Munari}, \& {Marrese}}]{2000A&A...354L..25T}
{Tomov}, T., {Munari}, U., \& {Marrese}, P.~M. 2000, \aap, 354, L25

\bibitem[{{Tomov} {et~al.}(2017){Tomov}, {Zamanov}, {Ga{\l}an}, \& {Pietrukowicz}}]{2017AcA....67..225T}
{Tomov}, T., {Zamanov}, R., {Ga{\l}an}, C., \& {Pietrukowicz}, P. 2017, \actaa, 67, 225

\bibitem[{{Tomov} {et~al.}(2016){Tomov}, {Stoyanov}, \& {Zamanov}}]{2016MNRAS.462.4435T}
{Tomov}, T.~V., {Stoyanov}, K.~A., \& {Zamanov}, R.~K. 2016, \mnras, 462, 4435

\bibitem[{{Tonry} {et~al.}(2018){Tonry}, {Denneau}, {Heinze}, {Stalder}, {Smith}, {Smartt}, {Stubbs}, {Weiland}, \& {Rest}}]{2018PASP..130f4505T}
{Tonry}, J.~L., {Denneau}, L., {Heinze}, A.~N., {et~al.} 2018, \pasp, 130, 064505

\end{thebibliography}
%

\begin{appendix}
\section{Optical spectrum of V4141~Sgr in 1984}\label{app:1984}
\begin{figure}[t]
\centering
\includegraphics[width=\columnwidth]{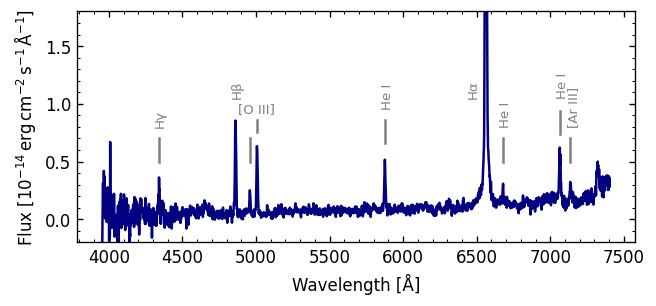}
\caption{Optical spectrum of V4141~Sgr from April 1984, with the positions of prominent emission lines indicated.}
\label{fig:1984}
\end{figure}

The spectrum of V4141~Sgr obtained in April 1984 \citep[see details in][]{1997A&A...327..191M} was first described by \citet{1987A&AS...68...51S}, who reported a possible detection of the Raman-scattered \ion{O}{vi} $\lambda$6830\,\AA{} line. However, this detection has not been confirmed in subsequent studies, including spectra obtained during quiescence. The spectrum in question is shown in Fig. \ref{fig:1984}, where Raman-scattered \ion{O}{vi} $\lambda$6830\,\AA{} and $\lambda$7088\,\AA{} lines are not readily visible. This is consistent with the absence of the \ion{He}{ii} $\lambda$4686\,\AA{} line in this dataset, as this feature is invariably present alongside Raman-scattered \ion{O}{vi} in symbiotic star spectra \citep[e.g.,][]{2019ApJS..240...21A}.

We also note that this spectroscopic dataset is listed in Table 2 of \citet{2025Galax..13....5K}, but it appears twice—once referenced to \citet{1987A&AS...68...51S} and once to \citet{1997A&A...327..191M}. Additionally, the observation date given for the latter reference is incorrect.

\section{Near-IR spectroscopy of~V4141~Sgr}\label{app:ir}
\begin{figure}
\centering
\includegraphics[width=\columnwidth]{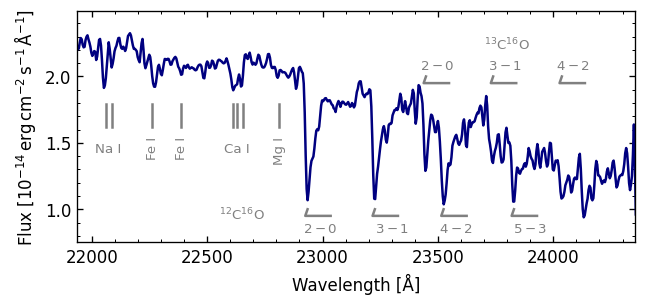}
\caption{K-band NTT spectrum of V4141~Sgr. The most prominent absorption features are labeled.}
\label{fig:ir_old}
\end{figure}

One of the key pieces of evidence supporting the presence of an M-type donor in V4141 Sgr—and ultimately leading to its classification as a confirmed symbiotic star by \citet{2000A&AS..146..407B}—was a previously unpublished near-IR K-band spectrum revealing an M6-type continuum. This spectrum was obtained on May 7, 1996, using the Infrared Spectrometer (IRSPEC) at the New Technology Telescope (NTT). More details on the observing program can be found in \citet{2003ASPC..303..163S}. The spectrum, shown in Fig. \ref{fig:ir_old}, confirms the presence of an M-type red giant, consistent with other classical symbiotic stars.

\section{Evolution of the hot component parameters}\label{app:hot}
\citet{2001A&A...376..978K} reported changes in the effective temperature of the "central star" during the observed brightness decline. We utilized published emission line fluxes of \ion{He}{ii} $\lambda$4868\,\AA{}, \ion{He}{i} $\lambda$5876\,\AA{}, and H$\beta$ to estimate the temperature and luminosity of the hot component at several epochs, namely, 1984 \citep[based on data from][]{1997A&A...327..191M}, 1987 \citep{1992PASP..104.1187G}, and 2002 \citep{2005A&A...435.1087L}, following the method outlined in \citet{1997A&A...327..191M}. The estimates assume a blackbody spectrum for the hot component and case B recombination for the emission lines. For these calculations, we adopted E$_{(B-V)}$ = 1.2 mag and a distance of d = 7 kpc \citep{1997A&A...327..191M}, consistent with a similar reddening value (E$_{(B-V)}$ = 1.12 mag) derived by \citet{2005A&A...435.1087L}. The distance estimate is roughly consistent also with the geometric ($\sim$7.4 kpc) and photogeometric ($\sim$6.0 kpc) distances of \citet{2021AJ....161..147B} based on the \textit{Gaia} EDR3 data.

In 1984, the temperature of the hot component was approximately 35--55 kK, with L$_h$ $\sim$ 1400 L$_\odot$. By 1987, temperature increased to $\sim$90 kK (L$_h$ $\sim$ 670 L$_\odot$) and remained similar in 2002 ($\sim$86 kK). During the 2025 outburst of V4141~Sgr, we assume that most of the hot component continuum emission is shifted to the optical (providing a lower limit on the luminosity if not). During the outburst, we estimate m$\rm _{bol}$ $\approx$ $V_{hot}\sim$13.4 mag. After correcting for reddening, $m_{\rm bol , 0} \approx 11.3$ mag, and the corresponding to an absolute bolometric magnitude of M$\rm _{bol}$ $\sim$ -4.5 mag and a luminosity of L$_h$ $\sim$ 5000 L$_\odot$. However, our most recent observations suggest that the brightness has not yet reached its maximum, meaning this value should be considered a lower limit for the luminosity during the outburst.

The evolution of V4141~Sgr during its recent outburst differs from that of classical symbiotic systems such as Z And, CI Cyg, and AX Per, whose outbursts are typically characterized by a decrease in temperature with little to no change in the luminosity of the hot component \citep[e.g.,][]{2010arXiv1011.5657M}. In contrast, the outbursts of AG Dra \citep{1995AJ....109.1289M} and Gaia18aen \citep{2020A&A...644A..49M} were accompanied by a luminosity increase by a factor of 5–10, suggesting that V4141~Sgr may share more similarities with these systems.

\end{appendix}
\end{document}